\documentclass[twocolumn,showkeys,preprintnumbers,amsmath,amssymb]{revtex4} 

\usepackage{graphicx}
\usepackage{dcolumn}
\usepackage{bm}
\usepackage{bbm}
\usepackage{amsmath}
\usepackage{mathtools}

\begin{document}

\title{Beyond linear coupling in microwave optomechanics}
\author{D.$~$Cattiaux$^{*}$,  X.$~$Zhou$^{**}$, S.$~$Kumar$^{*}$, I.$~$Golokolenov$^{*}$,  R.$~$R.$~$Gazizulin$^{*}$, A.$~$Luck$^{*}$, L.$~$Mercier de L\'epinay$^{***}$, M.$~$Sillanp\"a\"a$^{***}$, A.$~$D.$~$Armour$^{****}$, A.$~$Fefferman$^{*}$ and E.$~$Collin$^{*,\dag}$}

\address{(*) Univ. Grenoble Alpes, Institut N\'eel - CNRS UPR2940, 
25 rue des Martyrs, BP 166, 38042 Grenoble Cedex 9, France \\
          (**) IEMN, Univ. Lille - CNRS UMR8520, 
Av. Henri Poincar\'e, Villeneuve d'Ascq 59650, France \\
          (***) Departement of Applied Physics, Aalto University,  FI-00076 Aalto, Finland  \\
 	(****) Centre for the Mathematics and Theoretical Physics of Quantum Non-Equilibrium Systems and School of Physics and Astronomy, University of Nottingham, Nottingham NG7 2RD, United Kingdom 
 }

\date{\today}

\begin{abstract}
We explore the nonlinear dynamics of a cavity optomechanical system. Our realization 
consisting of a drumhead nano-electro-mechanical resonator (NEMS) coupled to a microwave cavity, allows for a nearly ideal platform to study the nonlinearities arising purely due to radiation-pressure physics. 
Experiments are performed under a strong microwave Stokes pumping which triggers mechanical self-sustained oscillations.
We analyze the results in the framework of an extended nonlinear optomechanical theory, and demonstrate that quadratic and cubic coupling terms in the opto-mechanical Hamiltonian have to be considered. 
Quantitative agreement with the measurements is obtained considering only genuine geometrical nonlinearities: no thermo-optical instabilities are observed, in contrast with laser-driven systems. 
Based on these results, we describe a method to quantify nonlinear properties of microwave optomechanical devices.
Such a technique, available now in the quantum electro-mechanics toolbox, but completely generic, is mandatory for the development of new schemes where higher-order coupling terms are proposed as a new resource, like Quantum Non-Demolition measurements, or in the search for new fundamental quantum signatures, like Quantum Gravity. 
We also find that the motion imprints a wide comb of extremely narrow peaks in the microwave output field, which could also be exploited in specific microwave-based measurements, potentially limited only by the quantum noise of the optical {\it and} the mechanical fields for a ground-state cooled NEMS device.  

\end{abstract}


\maketitle

%
\section{Introduction}
Combining mechanical resonators with dimensions of order a micron or less with superconducting circuit elements has led to an exciting field of research exploring the quantum properties of nanoelectromechanical systems (NEMS) \cite{pootzant}. Mechanical components can be efficiently coupled to superconducting qubits or integrated within optomechanical resonant cavities, providing a new resource for quantum device engineering 
\cite{cleland2010,lehnert2008,quantelecsimmonds,quantelec2}.
With a mechanical mode cooled to its quantum ground state, these NEMS circuits are also a new unique tool for fundamental experiments at the frontier of quantum mechanics \cite{ArmourBlencowe,sillanpaaintrique}.
 
In the context of quantum electronics, non-classical mechanical states can be used as a new support for quantum information storage and processing \cite{sillanpaaintrique,Delsing2019,clelandSAW,quantelecsimmonds,SlettenLehnert,sillanpaamultimode}.
By engineering the coupling between photons and phonons (i.e. {\it bath engineering}), non-reciprocal microwave quantum-limited on-chip components are being developed  \cite{clerknonrec,simmondsamplifnonrecip,kippenbergamplifnonrecip,jfink}; in order to couple quantum processors through optical photons, photon converters are being built around a quantum-mechanical degree of freedom \cite{convOskar,cindy}.

The capabilities offered by quantum NEMS devices are thus extremely rich, but are essentially all building on the {\it linear} parametric coupling between light and motion \cite{AKMreview}. For instance, using a pump properly detuned from the resonance cavity (i.e. {\it Anti-Stokes} sideband pumping), one can actively cool down a mechanical mode to its quantum ground state, or reversely  (i.e. {\it Stokes} pumping) amplify the mechanical motion \cite{sillanpaamultimode,schwabtones,Painter2,Painter3,delftsteeleOscill,Xzhou}. From two-tone schemes, one can devise back-action evading (BAE) measurements \cite{MikaBAE,SchwabBAE} that enable to beat the {\it standard quantum limit} by measuring one quadrature while feeding the back-action noise of the detector to the other one.

Beyond the standard optical frequency pulling proportional to mechanical position $x$, higher-order couplings appear to be also significant or even desirable for specific realizations:
from an $x^2$ coupling, one can measure the {\it energy} of the mechanics and build quantum non-demolition (QND) measurements \cite{Braginsky,clerkNEW}, i.e. measuring an eigenstate of the Hamiltonian while not perturbing its evolution. 
Successful experimental implementations of such nonlinear couplings have been realized using optics, with membrane-in-the-middle configurations \cite{Jayich} and {\it superfluid optomechanics} \cite{Childress}. 
On a more fundamental level, experiments which seek to use optomechanical systems to probe Gravity's quantum signatures,
 a completely new frontier of physics \cite{Vitali2015qgrav,AspelmeyerPlanck}, 
require necessarily to have characterized the higher order (beyond linear) mechanical couplings in the optomechanical Hamiltonian \cite{Qgravity}.

At very strong Stokes sideband pumping power, the mechanical mode enters {\it self-sustained oscillations} \cite{AKMreview}.
A rich multi-stable attractor diagram has been described theoretically \cite{Girvin1}, with specific phase-noise and amplitude-noise properties  \cite{Armour1,ArmourPRL,Tang1}. 
The mechanical amplitude of motion becoming very large, this coherent state dynamics is extremely sensitive to all nonlinearities present in the system; this had been discussed already in Ref. \cite{Dykman1}.
Experiments in this regime have been performed in the optical domain \cite{Favero1,Vahala1,Vahala2}; but the strong laser pump powers always produce dominant thermo-optical nonlinearities that require specific modeling \cite{Painter1}. 

In the present Article we report on experiments performed at low temperatures on a microwave optomechanical setup driven in the self-oscillating regime.
The dynamics imprints a comb in the microwave spectrum, from which we can measure more than 10 {\it 
extremely narrow} (width of order a few Hz) peaks.
Comb generation has been a revolution in optics \cite{Opticscombs}; it is thus natural to ask whether this effect could lead to a new technology.
Indeed, these combs could potentially be used for microwave-based readouts in quantum information processing (superconducting Qu-bits) and low temperature detectors (LTDs), where combs are usually synthesized for multiplexing purposes \cite{mazinLTD}. 
While efficient digital synthesizers are available today, routing all required signals down to millikelvin temperatures poses undeniable problems solved only by multiplying coaxial lines and generators.
Besides, while superconducting Qu-bits are not very sensitive to the quality of the GHz signals, digitally generated tones have a very poor phase noise at frequencies significantly offset from the carrier. Analog sources are the only currently foreseen possibility to generate low-phase noise GHz tones that are indispensable for many basic research applications - like e.g. optomechanics. 
For instance, an alternative technology proposed in the literature builds on the nonlinearity of superconducting quantum interference devices (SQUIDs) \cite{EricksonPRL,Irwin}, which enables to generate signals {\it inside} the cryostat. 
In comparison, our optomechanical combs are competitive: the peak width is of the same order (resolution of 1 part in $10^8$), the distribution of harmonics is much more homogeneous (equal spacing), and most importantly the amplitudes are very large (while the critical current of the SQUID junctions fixes a technological limitation).
One could thus imagine developing a new disruptive technology for {\it on-chip} synthesizing of harmonics, which could be a solution to the scaling problem faced today by cryogenic experiments. 
Furthermore, synchronization of several mechanical modes in the self-oscillation regime could further improve the phase noise performance of each individual mode
\cite{syncPRL}.
Ultimately, with a NEMS cooled to its ground state the stability of the output field would be limited only by quantum noise, and investigating these properties will be of fundamental interest \cite{ArmourPRL}.

\begin{figure}[t!]
		\centering
	\includegraphics[width=23cm]{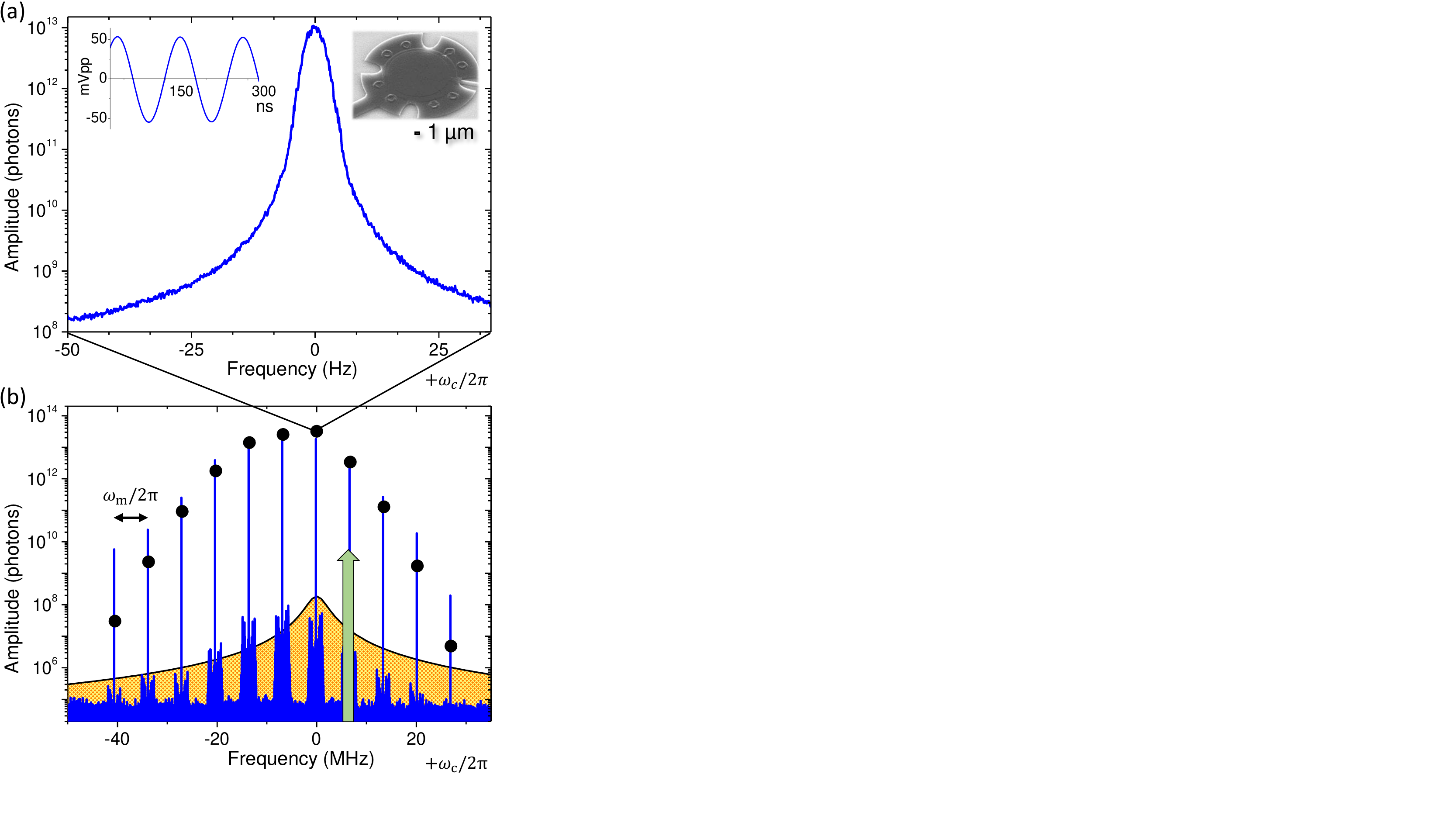}
	\vspace*{-0.5cm}
			\caption{
			(a) Main: Power spectral density (PSD, in units of photons of energy $\hbar \omega_c$) of the Stokes peak (i.e. at the frequency of the cavity) measured in the self-sustained oscillation regime at 214$~$mK (blue-detuned pump power $P_{in}$ of 6$~$nW with $\Delta=+\omega_m$). 
			In order to resolve the peak, the detection bandwidth was 0.2$~$Hz, for a span of $\pm 3.5~$kHz. 
			Left inset: Time domain measurement of the coherent signal (raw data units). Right inset: SEM picture of a drumhead type resonator. (b): PSD measurement of the comb produced by the strong applied power in same conditions (pump signal 6$~$nW, green arrow at its frequency $(\omega_c+\omega_{m})/2\pi$). The cavity (orange area) is displayed with an arbitrary amplitude and its linewidth $\kappa_{tot}/2\pi$ at scale. The black points are theoretical computation of the output amplitude of each measured peak (see text).}
			\label{fig_1}
\end{figure}

In the present work we demonstrate that the dynamics of the self-oscillating state is imprinted by {\it genuine geometrical nonlinearities} that can be fit, and we develop the full nonlinear theory giving the tools to extract nonlinear terms 
arising from the pure radiation-pressure coupling
up to the third order $\propto x^3$.
The agreement between experiment and theory is exceptional, and it gives us confidence in our level of understanding of the setup.
Building on these results, self-sustained oscillations in microwave optomechanical systems become a new tool enabling the experimental determination of the full nonlinear Hamiltonian at stake. This could be employed for instance in future quantum electronics circuits with specific schemes aiming at QND measurements \cite{Braginsky,clerkNEW}.

\section{Experiment}
\label{expt}

 We employ a standard microwave optomechanical system \cite{lehnert2008,lehnert2009} consisting of a microfabricated lumped microwave cavity resonator coupled to an aluminum drumhead  NEMS \cite{sillanpaaintrique} (see Fig. \ref{fig_1} a, right inset). The chip is installed into a commercial dilution cryostat with base temperature 10$~$mK, equipped with a high electron mobility transistor (HEMT) detection circuitry. The cryogenics, thermometry and measurement techniques have been described in Ref. \cite{Xzhou}. 

\begin{figure}[t!] \vspace*{-1cm}
		\centering
	\includegraphics[width=23cm]{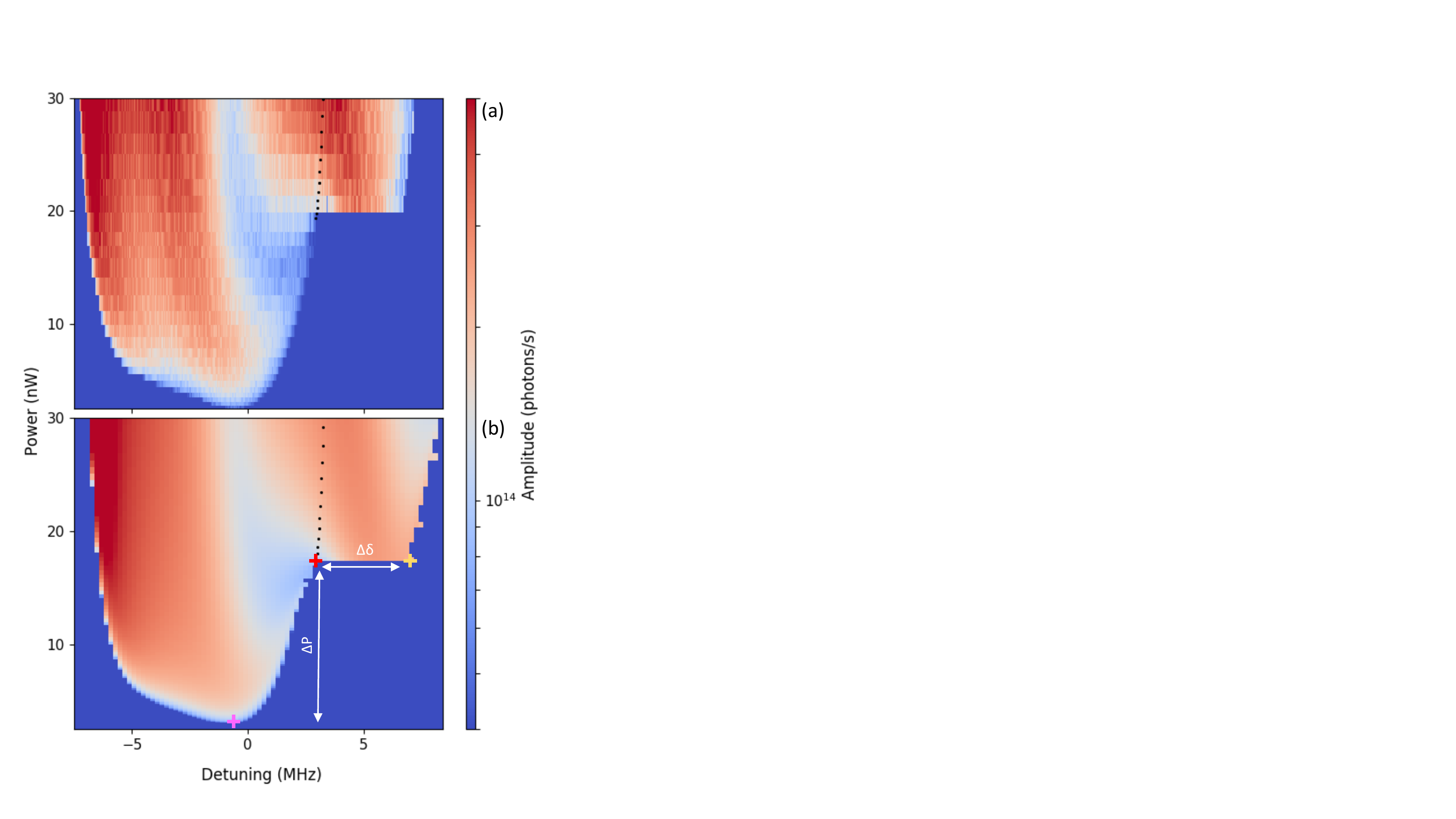}
	\vspace*{-0.5cm}
			\caption{
			(a): Measured output photon flux (Stokes peak at $\omega_c+\delta$) as a function of input power $P_{in}$ and detuning $\delta$ at 214$~$mK (pump tone at $\omega_c+\omega_m+\delta$). (b): Corresponding calculated colormap from the basic theory described in Refs. \cite{Girvin1,Armour1} (no nonlinearities, $g_1=g_2=0$ in Section \ref{theory}). The region on the right of the dashed line (high powers, positive detuning) is bistable and exists {\it only} when entering from the self-oscillating state (up-sweeping frequencies). The pink cross marks the minimum power necessary for self-oscillations, while the red cross corresponds to the position of the beginning of the hysteresis. The yellow cross marks the end of this bistable region (at same power). $\Delta P$ and $\Delta \delta$ are discussed in the text.}
			\label{fig_nononlin}
\end{figure}

 The chip is designed for reflection measurements. 
The aluminum microwave cavity resonates at  $\omega_c/2 \pi \approx 6.8~$GHz. The cavity displays a one-directional external coupling rate of $\kappa_{ext} /2 \pi\approx 2~$MHz, and a total damping rate of $\kappa_{tot} /2 \pi\approx 4~$MHz. We performed the experiment using the fundamental mode of the drum NEMS device which resonates around $\omega_m/2 \pi \approx 6.7~$MHz and exhibits a typical damping rate of about $\gamma_m/2 \pi\approx 150~$Hz at 50$~$mK. 
Details on the geometry and measured parameters can be found in Appendix \ref{characs}.
 
The optomechanical coupling mechanism arises from momentum transfer between light (i.e. photons) and mechanics (i.e. phonons). In the standard case of a Fabry-Perot cavity, the displacement of the movable end mirror (the mechanical degree of freedom) modulates the resonant frequency of the cavity (the optical mode).
Because of the retarded nature of the radiation pressure force when the laser light is detuned from the cavity frequency, this interaction gives rise to dynamical backaction allowing either active cooling or amplification of the mechanical motion.
In this respect, our experiment is analogous to optics but shifted in the microwave domain \cite{lehnert2008}; 
 the fundamental mode of the drumhead device corresponding to the movable mirror degree of freedom modulating the capacitance $C(x)$ of the electrical circuit \cite{AKMreview}.
 The Brownian motion of this mode then imprints sidebands in the microwave spectrum that we measure. 
 The motion amplitude being very small, no extra nonlinearity has to be considered and the optical damping (when cooling) and anti-damping (when amplifying) observed are linear in applied power $P_{in}$ \cite{AKMreview}.
This is used to calibrate the linear optomechanical interaction of our setup \cite{Xzhou}. We obtain a single photon-phonon coupling strength $\vert g_0\vert/2 \pi \approx 10~$Hz.

Blue-detuned pumping at $\omega_c+\Delta$ (with $\Delta > 0$) gives rise to downward scattering of photons, leading to the creation of phonons in the mechanical mode, hence enhancing the Stokes sideband. This is accompanied by a narrowing of the mechanical peak due to the antidamping backaction.
At very strong powers, the total mechanical damping can thus be totally canceled: this is called the parametric instability.
Above this threshold, the system enters into self-sustained oscillations, the amplitude of the mechanical motion being defined self-consistently  \cite{AKMreview}.
In this regime the mechanical amplitude of motion is so large (reaching several nanometers) that the mechanical sidebands are not limited to a couple of peaks: a full comb appears and can be measured (see Fig. \ref{fig_1} b). The peaks detected are not Lorentzian anymore, and their shape is defined by phase noise in the system \cite{Armour1}. 
They are extremely narrow (only a few Hz wide at GHz frequencies), essentially equally-spaced (by $\omega_m$) and of {\it extremely high amplitude}: they can even be detected without any HEMT pre-amplification.
As well, {\it all nonlinearities} in the device will impact this complex optomechanical dynamics.

At milliKelvin temperatures, 
heating arising from 
microwave absorption in dielectrics does not produce any thermal expansion: there are thus {\it no thermo-optical} nonlinearities in our system, in strong contrast with devices actuated by laser beams where they dominate \cite{Favero1,Vahala1,Vahala2,Painter1}. However, the strong pump signal required to reach the threshold of the parametric instability does give rise to heating effects.
This is carefully characterized and taken into account experimentally, see Appendix \ref{heatTLS}.
As such, the nonlinearities that prevail in microwave-based systems are 
of {\it geometrical} origin.

The experiment is performed in the mechanical self-induced oscillation regime, by measuring the output microwave signal corresponding to the Stokes peak (at frequency $\omega_{c}+\delta$), varying the detuning $\delta$ and the power $P_{in}$ of the input blue-detuned pump (at frequency $\omega_{c}+\omega_{m}+\delta$). 
The measured photon flux is shown in Fig. \ref{fig_nononlin} (a). 
It is obtained with a bandwidth of more than 10 times the width of the Stokes peak, such that the detected signal corresponds to the integrated PSD of the peak (as opposed to Fig. \ref{fig_1} which resolves its shape). 
As a comparison, the calculation based on the basic model of Refs. \cite{Girvin1,Armour1} is displayed in the bottom panel. 
The two plots are very similar, and display strikingly a {\it bistable region} at high powers and large positive detunings.  
However, calculation and theory {\it do not} match perfectly, which is expected: this has to be the signature of nonlinear effects which were neglected so far.

The region of the stability diagram which seems to be the most impacted by nonlinearities is precisely the hysteretic one (Fig. \ref{fig_nononlin} beyond the dashed line).
Therefore, in addition to the overall topography of the measured signal in the $(P_{in},\delta)$ space, we shall measure the importance of nonlinear features by 
reporting the position of the bistability in powers with respect to the beginning of the self-sustained region $\Delta P$, and its width in detuning $\Delta \delta$ (see Fig. \ref{fig_nononlin} b).

The question that arises is thus: which nonlinearities need to be included in a quantitative model?
One would immediately think about the {\it Duffing} effect in mechanical devices \cite{cross}, and correspondingly to the {\it Kerr} effect for the microwave cavity \cite{Maleeva}.
Both are discussed in Section \ref{results} and within the Appendices, but are {\it not} the dominant nonlinear features essentially because small frequency shifts have only a marginal impact on the optomechanical scheme itself.
This is in strong contrast with mechanics of directly-driven systems, where the Duffing nonlinearity generates a rich bistable dynamics, involving one or {\it even more} mechanical modes (through nonlinear mode-coupling)  \cite{nayfehbook,martial,roukesattractors,naturecoupl}.
In addition to geometrical nonlinearities, bottom-up devices have demonstrated a specific {\it material-dependent} feature: nonlinear damping
\cite{bachtoldnonlindamp}. While it could modify the dynamics if large enough, for top-down fabricated structures these effects are essentially negligible, even for high-amplitude motion of cantilever beams \cite{RSIcollin}; we discuss the point in Appendix \ref{heatTLS}.
We have thus to consider nonlinearities in the {\it coupling} itself, that is higher-order derivatives in the Taylor expansion of the coupling capacitance $C(x)$, which generate a modulation of the optomechanical interaction at harmonics of the mechanical resonance frequency. 

\section{Theory}	
\label{theory}

We investigate the influence of nonlinear position coupling on the dynamics of self-sustained oscillations. 
The wide separation of time-scales together with weak coupling and damping allows for a self-consistent approach in which the mechanical amplitude is slowly changing \cite{Girvin1,Armour1,Tang1,ArmourPRL,Dykman1}.
We start by writing a modified optomechanical Hamiltonian of the form (in the rotating frame of the drive optical field):
\begin{eqnarray}  \label{hamilton}
\hat{H} & = &-\hbar\left[ \Delta+g_{0}\left(\hat{b}+{\hat{b}}^{\dag}\right)  \right. \\  \nonumber
&+ & \left. \frac{g_{1}}{2}\left(\hat{b}+{\hat{b}}^{\dag}\right)^{2}+\frac{g_{2}}{2}\left(\hat{b}+{\hat{b}}^{\dag}\right)^{3}\right]{\hat{a}}^{\dag}\hat{a}\\ \nonumber 
& +& \hbar\omega_{m}{\hat{b}}^{\dag}\hat{b}+\hbar\Omega\left({\hat{a}}^{\dag}+\hat{a}\right)+\hat{H}_{\gamma} , 
\end{eqnarray}
where $\hat{a}$ and $\hat{b}$ are the photon and phonon annihilation operator, respectively. $g_{0} \propto d C/d x$ is the usual linear single photon-phonon coupling strength while we introduce $g_{1} \propto d^2 C/d x^2 $ and $g_{2}\propto d^3 C/d x^3$, respectively the quadratic and cubic coupling strengths. 
This expansion order is {\it necessary and sufficient} to provide quantitative fits of the data, see Section \ref{results}.
$C(x)$ is the cavity mode total capacitance while $x$ denotes the position collective degree of freedom of the first mechanical flexural mode (see Appendix \ref{characs} for details).
$\hat{H}_{\gamma}$ represents the external baths coupling Hamiltonian and $\Omega^2=\kappa_{ext} P_{in}/[\hbar (\omega_c+\Delta)]$ is the normalized driving term. In this case the equations of motion for both operators take the following form:
\begin{eqnarray}   \label{alphabeta}
\langle\dot{\hat{a}}\rangle &= &\left(\mathbbm{i}\Delta-\kappa_{tot}/2\right)\langle\hat{a}\rangle+\mathbbm{i}g_{0}\langle(\hat{b}+{\hat{b}}^{\dag})\hat{a}\rangle \\ \nonumber
&+&\mathbbm{i}\frac{g_{1}}{2}\langle(\hat{b}+{\hat{b}}^{\dag})^{2}\hat{a}\rangle+\mathbbm{i}\frac{g_{2}}{2}\langle(\hat{b}+{\hat{b}}^{\dag})^{3}\hat{a}\rangle-\mathbbm{i}\Omega , \\
\langle\dot{\hat{b}}\rangle &=&-\left(\mathbbm{i}\omega_{m}+\gamma_{m}/2\right)\langle\hat{b}\rangle+\mathbbm{i}g_{0}\langle{\hat{a}}^{\dag}\hat{a}\rangle\\ \nonumber
&+& \mathbbm{i}g_{1}\langle(\hat{b}+{\hat{b}}^{\dag})\hat{a}^{\dag}\hat{a}\rangle+ \mathbbm{i}\frac{3g_{2}}{2}\langle(\hat{b}+{\hat{b}}^{\dag})^{2}\hat{a}^{\dag}\hat{a}\rangle . 
\end{eqnarray}
When the amplitudes of both fields are large enough, we can neglect quantum fluctuations and use the standard semiclassical approach.
We write for both the optics $\langle\hat{a}\rangle\to\alpha$ and the mechanics $\langle\hat{b}\rangle\to\beta$, leading to:
\begin{eqnarray}
\dot{\alpha} &=& \left(\mathbbm{i}\Delta-\kappa_{tot}/2\right)\alpha+\mathbbm{i}g_{0}(\beta+\beta^{*})\alpha\\ \nonumber
&+& \mathbbm{i}\frac{g_{1}}{2}(\beta+\beta^{*})^{2}\alpha+\mathbbm{i}\frac{g_{2}}{2}(\beta+\beta^{*})^{3}\alpha-\mathbbm{i}\Omega ,\\ \label{beta}
\dot{\beta} &=& -\left(\mathbbm{i}\omega_{m}+\gamma_{m}/2\right)\beta+\mathbbm{i}g_{0}\vert\alpha\vert^{2}\\ \nonumber 
&+& \mathbbm{i}g_{1}(\beta+\beta^{*})\vert\alpha\vert^{2}+\mathbbm{i}\frac{3g_{2}}{2}(\beta+\beta^{*})^{2}\vert\alpha\vert^{2} . \label{beta2}
\end{eqnarray}
This system of coupled equations can be solved by means of the ansatz for $\beta$ \cite{Armour1}:
\begin{equation} \label{betac}
\beta=\beta_{c}+Be^{-\mathbbm{i}\phi} e^{-\mathbbm{i}\omega t},
\end{equation}
$\beta_{c}$ being related to a static deflection $x_c$ of the drum, and $B  e^{-\mathbbm{i}\phi}$ corresponding to the (complex valued) coherent motion.
In the following, we shall neglect the $\beta_{c}$ term; it is indeed responsible only for a tiny frequency shift of the mechanical resonance, which impacts only marginally the thought limit cycle state. For the same reason we did not include the mechanical (Duffing) nonlinearity in Eq. (\ref{hamilton}), see Appendix \ref{duffing} for a detailed discussion on these issues.

For convenience, we now introduce a shifted detuning $\Delta'=\Delta+g_{1}B^{2}$, and two renormalized coupling parameters $G=2g_{0}+3B^{2}g_{2}$ and $\bar{G}=2g_{0} +6 B^{2}g_{2}$. 
The optical amplitude equation takes now the form:
\begin{eqnarray}
\dot{\alpha}&=&\left[\mathbbm{i}\Delta'-\kappa_{tot}/2+\mathbbm{i}GB\cos(\omega t+\phi)\right. \\ \nonumber
&&\left.+\mathbbm{i}g_{1}B^{2}\cos(2\omega t+2\phi) \right. \\ \nonumber 
&& \left. +\mathbbm{i}g_{2}B^{3}\cos(3\omega t+3\phi)\right] \alpha-\mathbbm{i}\Omega .
\end{eqnarray}
The solution can be found via the mathematical transform 
 described in Ref. \cite{Girvin1}. We define $\tilde{\alpha} = \alpha e^{\mathbbm{i}\Theta}$ with:
\begin{eqnarray}
\Theta (t) & = & -\frac{G B}{\omega}\sin(\omega t +\phi)-\frac{g_{1}B^{2}}{2\omega}\sin(2\omega t+2\phi) \\ \nonumber
& -&\frac{g_{2}B^{3}}{3\omega}\sin(3\omega t+3\phi),
\end{eqnarray}
leading to the simpler dynamics equation:
\begin{equation}  \label{alphatilde}
\dot{\tilde{\alpha}}=\left(\mathbbm{i}\Delta'-\frac{\kappa_{tot}}{2}\right)\tilde{\alpha}-\mathbbm{i}\Omega e^{\mathbbm{i}\Theta}.
\end{equation}
We now use the Jacobi-Anger expansion three times (on the three terms defining $\Theta$):
\begin{equation}
 f(t)=-\mathbbm{i}\Omega e^{\mathbbm{i}\Theta (t)}=-\mathbbm{i}\Omega\sum_{n\in\mathbb{Z}}f_{n}e^{\mathbbm{i}n(\omega t+\phi)}
\end{equation}
where:
\begin{eqnarray}
\label{Jn}
f_{n}&=& \sum_{m\in\mathbb{Z}}\sum_{p\in\mathbb{Z}}(-1)^{m+p}  \\ \nonumber
 &\times & J_{p}\left(\frac{-g_{2}B^{3}}{3\omega}\right)J_{m}\left(\frac{-g_{1}B^{2}}{2\omega}\right) J_{3p+2m+n}\left(\frac{-GB}{\omega}\right) .
\end{eqnarray}
Here, $J_{n}$ is the Bessel function of the first kind. 
 Fourier transforming Eq. (\ref{alphatilde}), we write $\tilde{\alpha}(t)=\sum_{n \in \mathbb{Z}} \tilde{\alpha}_n e^{\mathbbm{i}n \omega t}$ with:
\begin{equation}
\tilde{\alpha}_{n}= \frac{-\mathbbm{i}\Omega e^{\mathbbm{i}n\phi}f_{n}}{\mathbbm{i}\left(n\omega-\Delta' \right)+\kappa_{tot}/2},
\end{equation}
and hence:
\begin{eqnarray}
\vert\alpha\vert^{2}&=&\vert\tilde{\alpha}\vert^{2}=\sum_{(n,n')\in\mathbb{Z}^2}\tilde{\alpha}_{n}\tilde{\alpha}_{n'}^{*}e^{\mathbbm{i}(n-n')\omega t}  \\ \nonumber 
&=&\sum_{(n,n')\in\mathbb{Z}^2}  \Omega^{2} \frac{e^{\mathbbm{i}(n-n')\omega t} e^{\mathbbm{i}(n-n')\phi} f_{n}f_{n'}}{h_{n}h_{n'}^{*}} \\ \nonumber
&=& \sum_{q\in\mathbb{Z}} e^{-\mathbbm{i}q \omega t}  e^{-\mathbbm{i}q\phi} \,  \Omega^{2}  \left[ \sum_{n \in\mathbb{Z}} \frac{f_{n}f_{n+q}}{h_{n}h_{n+q}^{*}} \right]  \\ \label{comb}
& = &  \sum_{q\in\mathbb{Z}} e^{-\mathbbm{i}q \omega t} \eta_q , \label{etaq}
\end{eqnarray}
where $h_{n}= \mathbbm{i}(n\omega-\Delta')+\kappa_{tot}/2$, and we used the change of variable $q=n'-n$ in the penultimate line.

\begin{figure}[t!]
		\centering\offinterlineskip
	\hspace*{-0.25cm}
	\includegraphics[trim=5cm 2cm 0cm 4cm,clip=true,width=9.7cm]{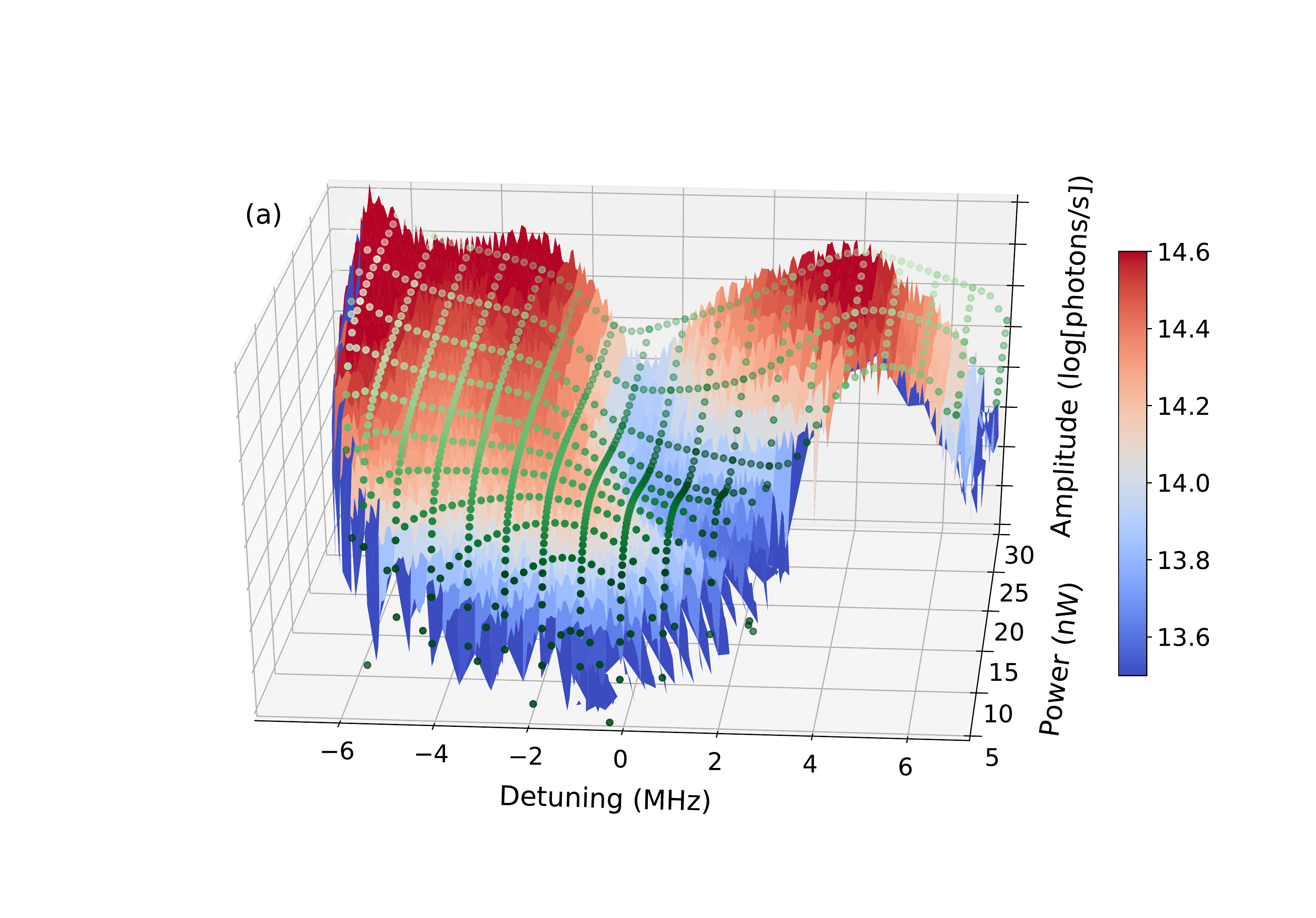}
	\hspace*{-0.25cm}
	\includegraphics[trim=5cm 0cm 0cm 4cm,clip=true,width=9.7cm]{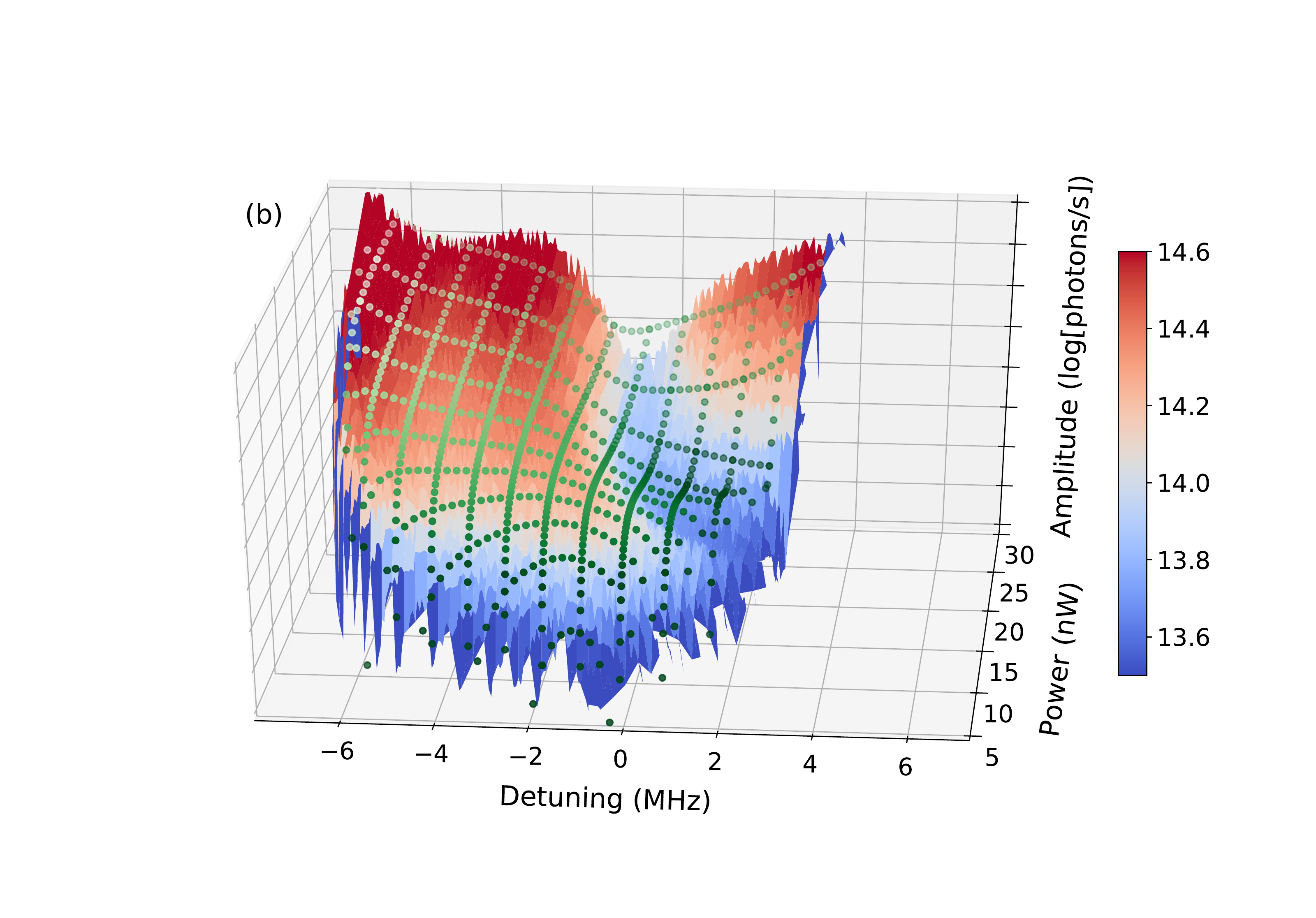}
	\vspace*{-0.5cm}
			\caption{
			(a): Output photon flux of the self-oscillating (Stokes) peak at frequency $\omega_{c}+\delta$, as a function of both the power $P_{in}$ and the detuning $\delta$ of the input pump signal (pump frequency $\omega_{c}+\omega_{m}+\delta$, with -7 MHz$<\delta<$+7 MHz) at 214$~$mK. The colormap is experimental data measured up-sweeping both the pump detuning (from $\delta = -7$ MHz to $\delta = +7$ MHz) and the pump power, and green points are theoretical fits computed by solving self-consistently Eq. (\ref{equatosolve}), $\gamma_m+\gamma_{BA}=0$, see text. (b): Experimental colormap measured down-sweeping the pump detuning (from $\delta = +7$ MHz to $\delta = -7$ MHz) with pump power swept upwards. Green points are also theoretical computations; the hysteresis of the large power and large detuning region is clearly visible. }
			\label{fig_2}
\end{figure}

Inserting Eq. (\ref{etaq}) into Eq. (\ref{beta}) and preserving only terms oscillating at $-\omega$ (rotating wave approximation),
we obtain the dynamics equation for the mechanics:
\begin{eqnarray}
\dot{\beta}&=&-\left(\mathbbm{i}\omega_m+\frac{\gamma_{m}}{2}\right)\beta+\mathbbm{i}\frac{\bar{G}}{2}\eta_{1} e^{-\mathbbm{i}\omega t} \\ \nonumber
&+&\mathbbm{i}g_{1}\eta_{0}\beta+\mathbbm{i}g_{1}\eta_{2}\beta^{*}e^{-2\mathbbm{i}\omega t} \\ \nonumber 
&+& \mathbbm{i}\frac{3}{2}g_{2}\eta_{-1}\beta^{2}e^{\mathbbm{i}\omega t}+\mathbbm{i}\frac{3}{2}g_{2}\eta_{3}(\beta^{*})^{2}e^{-3\mathbbm{i}\omega t}.
\end{eqnarray}
We can now recast this expression introducing the optical backaction terms, namely the 
optical spring term $\delta\omega$ and the damping term $\gamma_{BA}$:
\begin{equation}
\dot{\beta}=-\left[\mathbbm{i}\omega+\frac{1}{2}(\gamma_{m}+\gamma_{BA})\right]\beta, \label{equanonlin2}
\end{equation}
with $\omega=\omega_{m}+\delta\omega$ now explicitly defined, and:
\begin{eqnarray}
\gamma_{BA} &= &-2\Re[X],\\
\delta\omega &= &-\Im[X],
\end{eqnarray}
where $X$ is written as:
\begin{eqnarray}
X &=& \mathbbm{i}\Omega^{2}\left[\frac{\bar{G}}{2B}\sum_{n \in\mathbb{Z}}\frac{f_{n}f_{n+1}}{h_{n}h_{n+1}^{*}} \right. \\ \nonumber 
&+& \left.  g_{1}\sum_{n\in\mathbb{Z}}\frac{f_{n}^{2}}{\vert h_{n}\vert^{2}}+g_{1}\sum_{n\in\mathbb{Z}}\frac{f_{n}f_{n+2}}{h_{n}h_{n+2}^{*}}\right.\\ \nonumber
&+& \left.\frac{3g_{2}B}{2}\sum_{n\in\mathbb{Z}}\frac{f_{n}f_{n+1}}{h_{n+1}h_{n}^{*}}+\frac{3g_{2}B}{2}\sum_{n\in\mathbb{Z}}\frac{f_{n}f_{n+3}}{h_{n}h_{n+3}^{*}}\right].
\end{eqnarray}
One can thus find all the stable states by solving self-consistently the equation which cancels the effective damping $\gamma_{m}+\gamma_{BA}$, ensuring that:
\begin{equation}
 \dot{B} =0. \label{equatosolve}
\end{equation}
In practice, it is sufficient to solve the limit-cycle equation neglecting all kinds of mechanical shifts, assuming $\omega=\omega_m$ in Eq. (\ref{Jn}).
Details on the self-consistent determination of optomechanical stable states can be found in Appendix \ref{stablestates}.

Following the same procedure as for $\tilde{\alpha}$, the optical field amplitude in the cavity takes the form  $\alpha = \sum_{n \in \mathbb{Z}} \alpha_n e^{\mathbbm{i}n \omega t}$ with:
\begin{equation}
 \alpha_n =  \sum_{q \in \mathbb{Z}} \frac{-\mathbbm{i}\Omega e^{\mathbbm{i}n\phi}f_{q} f_{q-n}}{h_q}.
\end{equation}
This expression highlights the fact that the optomechanical coupling imprints a {\it comb structure} in the photon field.  
We can then compute the output photon flux  $\dot{N}_{out,n}$ of each comb peak $n$ as:
\begin{equation}\label{Ndot}
\dot{N}_{out,n} = \kappa_{ext} \vert \alpha_n\vert^2, 
\end{equation}
where we made use of the well-known input-output relation linking intra-cavity fields and output traveling fields \cite{devoret}.
$n=0$ corresponds to the pump tone at frequency $\omega_c+\omega_m$, $n=-1$ to the Stokes sideband at $\omega_c$ and $n=1$ to the anti-Stokes sideband at $\omega_c+2\omega_m$ (see black dots in  Fig. \ref{fig_1} b).

\section{Role of geometric nonlinearities}	
\label{results}

The aim is thus now to go beyond Fig. \ref{fig_nononlin}, and obtain {\it quantitative agreement} between theory and experiment.
The theory in the previous section allows us to calculate the amplitude of the mechanical motion including geometrical nonlinearities in the couplings. However, to obtain estimates of the mechanical frequency we need to also include important contributions from other effects, especially the Duffing nonlinearity of the drum. 

\begin{figure}[t!]
		\centering\offinterlineskip
	\hspace*{-0.25cm}
	\includegraphics[trim=5cm 2cm 0cm 4cm,clip=true,width=9.7cm]{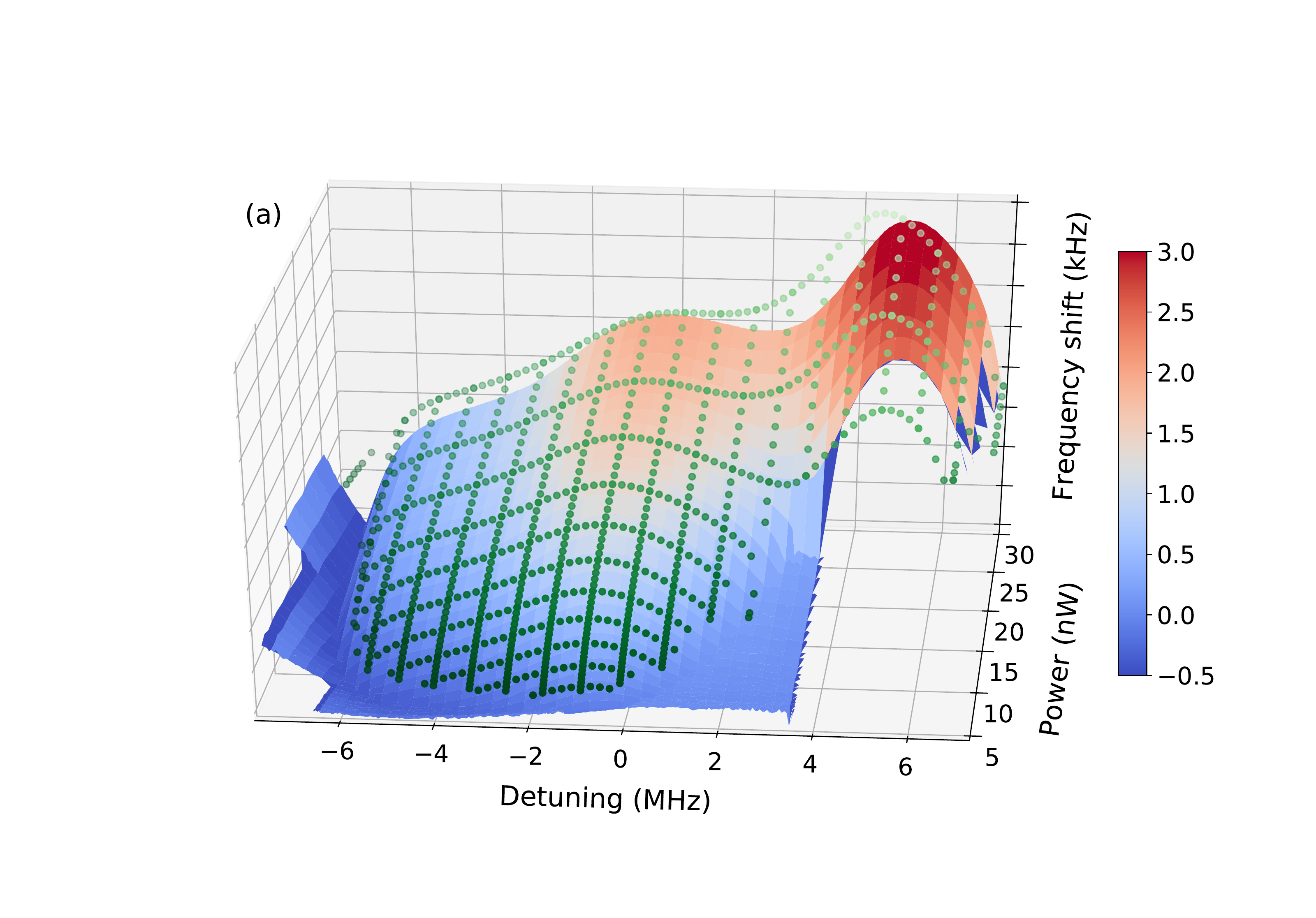}
	\hspace*{-0.25cm}
	\includegraphics[trim=5cm 0cm 0cm 4cm,clip=true,width=9.7cm]{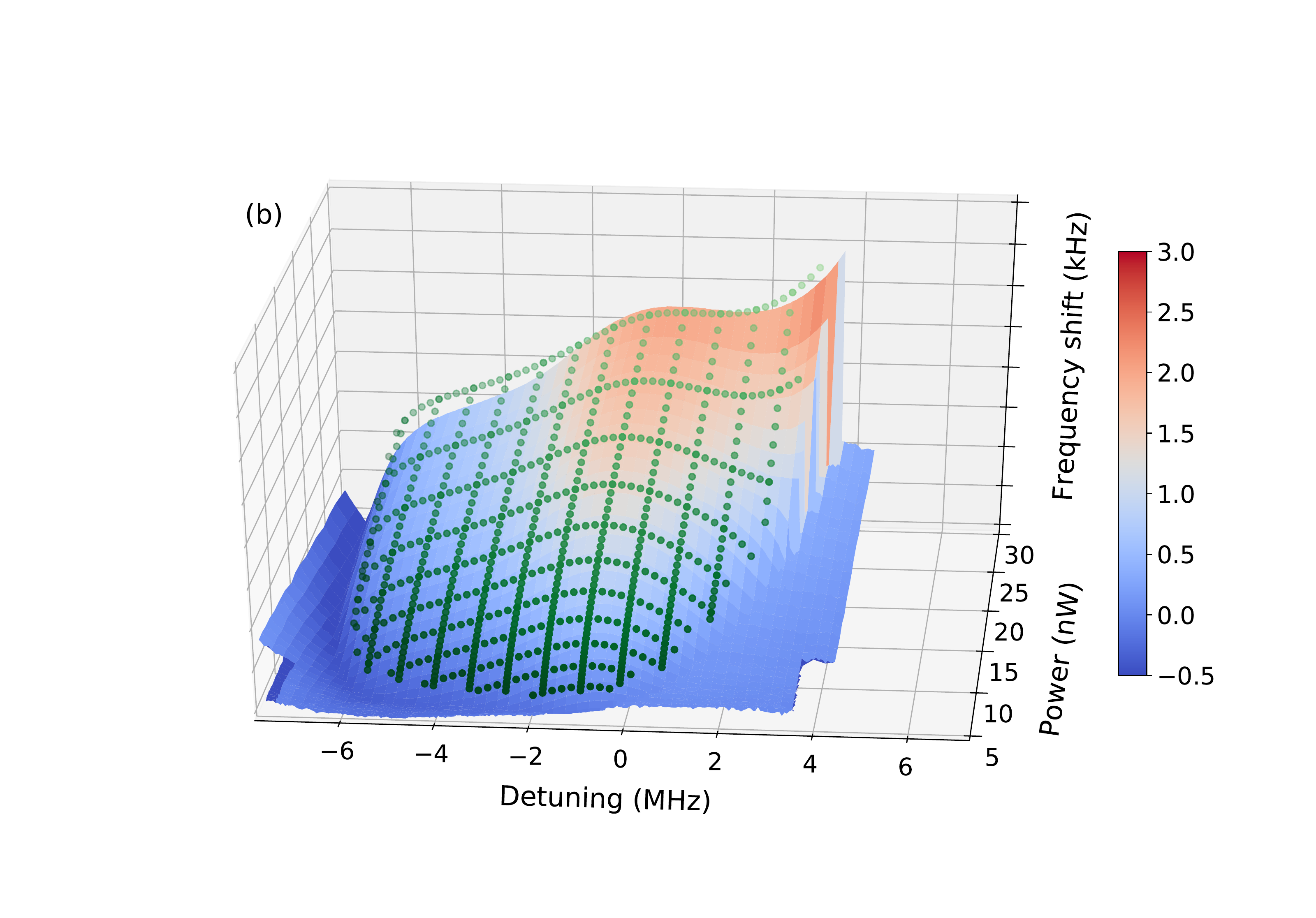}
	\vspace*{-0.5cm}
			\caption{
			(a): Mechanical frequency shift of the self-oscillating (Stokes) peak as a function of both the power $P_{in}$ and the detuning $\delta$ of the input pump signal (same conditions as Fig. \ref{fig_2}, sweeping $\delta$ towards positive values). Note that in the hysteretic region, the calculated points lie slightly below the experimental ones, but obviously match the threshold position of Fig. \ref{fig_2}.  (b): Experimental colormap measured down sweeping the pump detuning. Green points are theoretical computations, see text.}
			\label{fig_3}
\end{figure}

As soon as the system self-oscillates, the actual cavity frequency is slightly renormalized in $\omega_c'=\omega_c - g_1 B^2$.
Besides, there is also a material-dependent shift with a logarithmic power-dependence that is attributed to Two-Level-Systems present in the dielectrics \cite{TLS}, which is taken into account (Appendix \ref{heatTLS}). On the other hand, the cavity Kerr nonlinearity $\xi_c$ is expected to be extremely small for our device \cite{paperAlessandro,Maleeva}; we give an upper bound in Tab. \ref{tab_1}, see discussions in Appendices \ref{heatTLS} and \ref{duffing} for details.
The mechanical resonance is also renormalized by the optomechanical coupling, with a tiny frequency shift $\delta \omega$ (see Section \ref{theory}).
However, the dominant source of mechanical frequency shift is due to the Duffing effect (i.e. the mechanical nonlinearity arising from the stretching of the drum \cite{cattiaux}), leading to $\omega_m'=\omega_m+\delta \omega+\xi_m  B^2$, with a normalized Duffing parameter $\xi_m$ in Hz per phonon.
For simplicity, we will omit the prime on $\omega_c$ and $\omega_m$, remembering in the following that the measured mechanical frequency shift includes all terms.

The measured output photon flux is plotted in Fig. \ref{fig_2} as a function of detuning $\delta$ and power $P_{in}$ (same data as Fig. \ref{fig_nononlin} top panel, 214$~$mK).
The amplitude of the signal is extremely large, but the most striking feature is the bistable region at high powers and positive detunings.
The measured mechanical frequency shift is shown in Fig. \ref{fig_3}; strikingly, we find that it is largest in the bistable regime. 

This mechanical shift cannot be captured by the optomechanical contribution $\delta \omega$ alone. One has to take into account the Duffing effect to quantitatively fit it (see below).
However, the mechanical frequency shifts remain very small (a few kHz at most, see Fig. \ref{fig_3}); we thus verified that they have only a marginal impact on the limit cycle dynamics (i.e. the amplitudes, $B$), see Appendix \ref{duffing}.

In the hysteretic region the amplitude $B$ becomes very large, hence the optomechanical response becomes sensitive to the nonlinear coupling coefficients, $g_1$ and $g_2$. 
For symmetry reasons (see Appendix \ref{characs}), the sign of the $g_0$ parameter is irrelevant and we 
take it to be positive for simplicity.
However then, the sign of the other coefficients is uniquely defined. 

\begin{figure}[t!] \vspace*{-1cm}
		\centering
	\includegraphics[width=9.5cm]{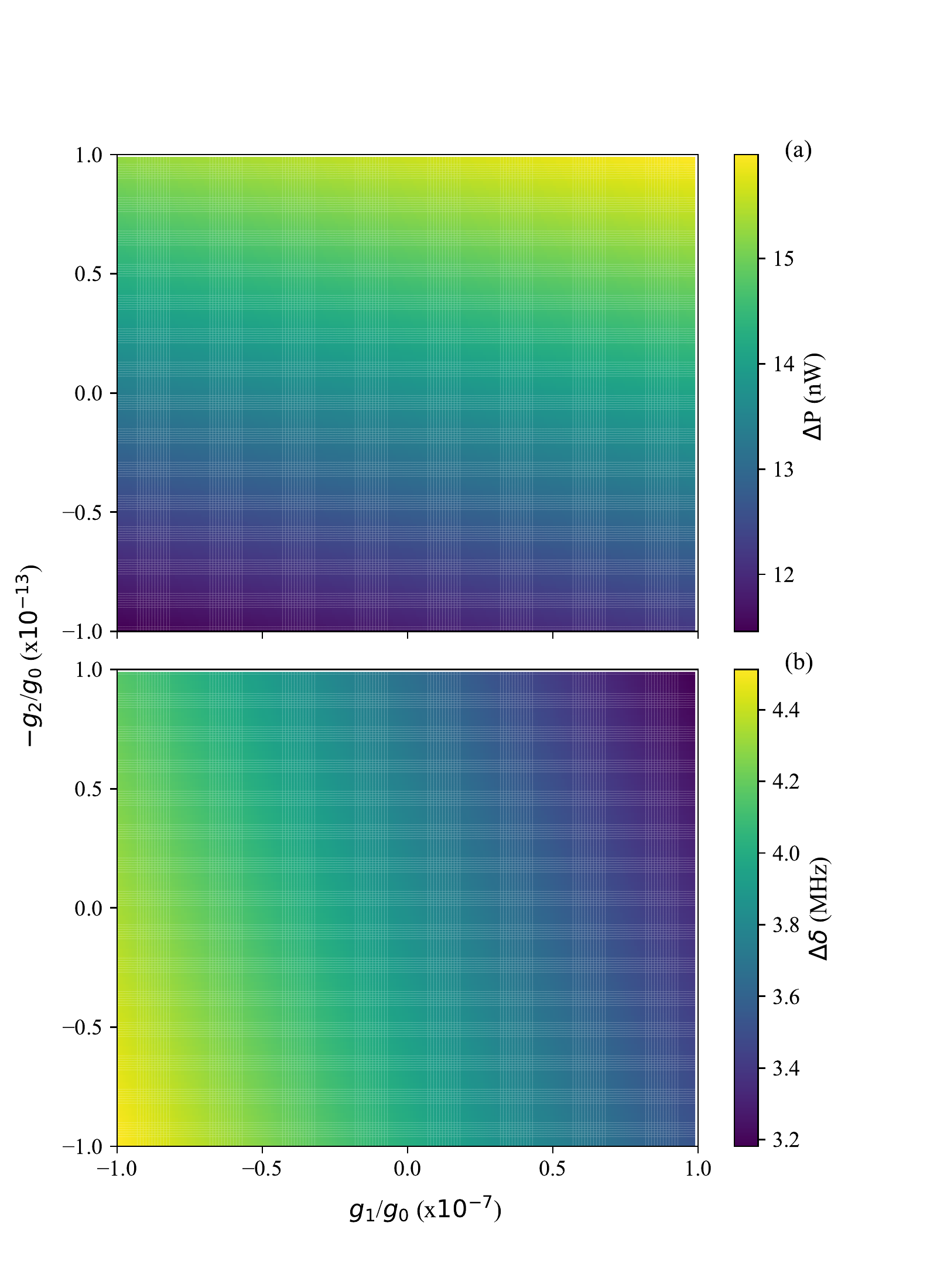}
	\vspace*{-0.5cm}
			\caption{
			(a): Calculated $\Delta P$ parameter as a function of $g_1,g_2$ coefficients. (b): Calculated $\Delta \delta$ parameter as a function of $g_1,g_2$. Both are essentially described by plane equations, with each nonlinear coefficient being the leading one for one of the parameters ($g_1$ for $ \Delta \delta$ and $g_2$ for $ \Delta P$, see text). Full colormaps are also presented in a matrix form in Appendix \ref{fitting}, Fig. \ref{fig_5}.}
			\label{fig_g1g2}
\end{figure}

To calculate the amplitudes of the limit cycles (and hence the photon flux) using the approach in Sec. \ref{theory}, the only free parameters are
the quadratic and cubic nonlinear coupling terms $g_1$ and $g_2$, respectively. 
These two coefficients have a different impact on the calculated flux: around our best fit parameters, $g_1$ narrows/broadens the self-oscillating region with respect to detuning (altering the $\Delta \delta$ parameter), while $g_2$ mostly shifts the bistable feature to higher/lower powers ($\Delta P$ parameter). 
This is represented in Fig. \ref{fig_g1g2}; in Appendix \ref{fitting}, full colormaps calculated for different nonlinear parameters are also displayed (Fig. \ref{fig_5}).
Indeed at the same time the overall {\it shape} of the theoretical maps displayed in Fig. \ref{fig_2} (flux) and Fig. \ref{fig_3}  (mechanical frequency) are very sensitive to the nonlinear parameters. 
We can therefore reasonably well determine the values of these two terms, typically within a factor of 2 (see Appendix \ref{fitting}).
The theoretical fits are displayed as green dots in Figs. \ref{fig_2} and \ref{fig_3}; as a comparison the colormap of Fig. \ref{fig_nononlin} bottom panel is computed for $g_1=g_2=0$.

We performed this procedure at various cryostat temperatures. However, because of microwave absorption in the materials, the drum temperature is not homogeneous over the complete measured range of $(\delta,P_{in})$. This effect is taken into account, see Appendix \ref{heatTLS}. 
The most constrained point for the definition of the couple $(g_1,g_2)$ is the junction between the main stable region and the bistable part, defined by the red cross mark in Fig. \ref{fig_nononlin}. We shall thus define an effective temperature $T_{ef\!f}$ characteristic of the fit at this precise point.

\begin{table}[!ht]
\begin{center}
\hspace*{-0.4cm}
  \begin{tabular}{ |c|c|c|c|c|c|}
    \hline
    $T$(mK) & $T_{ef\!f}$(mK) & $g_{1}/g_{0}$ & $g_{2}/g_{0}$ & $\xi_m$(Hz)    & $\xi_c$(Hz) \\ \hline
   cryo.          &  $\pm 20~\%$    & within $\times 2$ &   within $\times 2$ &  $\pm 10~\%$ & est. \\ \hline
    417 & 520 & $+1.\times10^{-7}$ & $-10.\times10^{-14}$ & $+2.1\times10^{-9}$  & $-10^{-4}$\\ \hline
    215 & 320 &   idem    &  idem  &  idem   &  idem \\ \hline
    50  & 290 &   idem    &  idem  &  idem   &  idem \\ 
    \hline
  \end{tabular}
      \caption{Fitted parameters at different temperatures. ``cryo.'' is the cryostat measured temperature, while $T_{ef\!f}$ is the characteristic fit temperature; nonlinear couplings are given in units of $g_0$, with $g_0>0$. The Kerr parameter of the cavity is estimated  (see text). } \label{tab_1}
\end{center}
\end{table}

From the measured mechanical frequency shift (Fig. \ref{fig_3}), we can finally fit the Duffing term $\xi_m$. 
The summary of our results is given in Tab. \ref{tab_1}.
Within our error bars, we can infer a {\it unique set} of parameters that fits all temperatures. This is a strong evidence that the nonlinear features $g_1, g_2$ and $\xi_m$ are of {\it geometrical} origin. We give in Appendix \ref{characs} theoretical estimates obtained from basic arguments: a circular plate stretching nonlinearity for $\xi_m$ \cite{cattiaux} and a corresponding plate-capacitor nonlinear expansion for $g_1, g_2$ \cite{feynmanbook}.
The magnitudes match our findings within typically a factor of 2, apart from $g_2$ whose prediction is the worst because of the crudeness of the plate capacitor analytic expansion.

\section{Conclusion}

We report on microwave optomechanical experiments performed in the self-sustained oscillation regime.
The output spectrum of a microwave cavity resonating around 6.8$~$GHz coupled to a 6.7$~$MHz drumhead mechanical device is measured as a function of input power $P_{in}$, pump frequency detuning $\delta$ and temperature. A high amplitude and narrow-peak comb structure is measured in the output spectrum, and fit to theory.

We demonstrate that the limit cycle dynamics is sensitive to  nonlinearities in the optomechanical coupling.
We therefore present a theory that goes beyond the standard linear optomechanical Hamiltonian, introducing quadratic and cubic terms $g_1$ and $g_2$. 
Data is fit quantitatively, and we show that these $g_1$ and $g_2$ must be of geometrical origin, as opposed to the thermo-optical nonlinear features present in laser driven systems.

The work described here can thus be proposed as a new method to characterize nonlinearities in microwave nanomechanical platforms.
With the development of new quantum-limited optomechanical schemes building on higher-order couplings \cite{clerkNEW,Qgravity}, it represents a very useful new resource. 
The method is also particularly straightforward since it simply relies on the strong pumping of the mechanics via the microwave field.
Besides, the generated comb itself could be used in schemes requiring microwave multiplexing. One could imagine specific designs with multiple cavities and NEMS producing much wider combs; 
adding DC gates would also enable frequency tuning \cite{roukesJAP}.

Finally, we note that our microwave coupled mechanical devices are fully compatible with ultra-low temperature cryostats capable of operating below 1$~$mK \cite{Xzhou}. Assuming equilibration of devices like the one used here can be achieved under such conditions, then they will naturally operate within the quantum regime. Further work will be needed to understand the extent to which the nonlinear coupling terms will squeeze the quantum fluctuations leading to an amplitude of motion that is more precisely defined than that of a coherent state \cite{ArmourPRL}. Furthermore, measuring the rate at which the system switches between the co-existing dynamical states that arise in the nonlinear regime when the system is in the quantum regime will provide important new insights into fundamental processes such as quantum activation \cite{QActiveDyk,activeDyk}. 

\vspace*{1cm}
(\dag) Corresponding Author: eddy.collin@neel.cnrs.fr

\begin{acknowledgements}

We acknowledge the use of the N\'eel {\it Cryogenics} facility with especially Anne Gerardin for realization of mechanical elements. 
E.C. would like to thank  I. Favero, A. Monfardini, F. Levy-Bertrand and M. Dykman for very useful discussions.
We acknowledge support from the ERC CoG grant ULT-NEMS No. 647917 (E.C.), StG grant UNIGLASS No. 714692 (A.F.), the STaRS-MOC project from {\it R\'egion Hauts-de-France} and ISITE-MOST project (X.Z.). A.D.A. was supported through a Leverhulme Trust Research Project Grant (RPG-2018-213), and M.S. was supported by the Academy of Finland (contracts 308290, 307757, 312057), by the European Research Council (615755-CAVITYQPD), and by the Aalto Centre for Quantum Engineering. The work was performed as part of the Academy of Finland Centre of Excellence program (project 312057). We acknowledge funding from the European Union's Horizon 2020 research and innovation program under grant agreement No.~732894 (FETPRO HOT).
The research leading to these results has received funding from the European Union's Horizon 2020 Research and Innovation Programme, under grant agreement No. 824109, the European Microkelvin Platform (EMP).

\end{acknowledgements}

\appendix

\section{Drumhead characteristics}
\label{characs}

The mechanical device used in this work is a typical aluminum drumhead \cite{sillanpaaintrique}.
As can be seen on the SEM picture in Fig. \ref{fig_1}, the actual structure is rather complex; we will simply approximate it as two discs of radius $R$ (one being fixed and the other movable) separated by a gap $d$. The thickness of the drum is $e$. These geometrical characteristics are summarized in Tab. \ref{tab_2} together with typical material parameters. 

\begin{table}[!t]
\begin{center}
  \begin{tabular}{ |c|c|c|c|c|c|}
    \hline
    $R$ (nm) & $d$ (nm) & $e$ (nm) & $E$ (GPa) & $\rho$ (kg/m$^3$)    & $\nu$ \\ \hline
   est.     &   est.    &  est. &   bulk val.&   bulk val. &  bulk val. \\ \hline
   $8\,500$   & 150 & $170$ & $70$ & $2\,700$  & $0.35$\\ 
    \hline
  \end{tabular}
      \caption{Typical drumhead NEMS parameters; the in-built stress is estimated to be $<60$ MPa  (see text). 
      Corresponding mode effective mass $m_{ef\!f}=2.3\times 10^{-14}~$kg and spring constant $k_{ef\!f}=41.~$N/m.} \label{tab_2}
\end{center}
\end{table}

These numbers are estimated from the Kirchhoff-Love theory of plates, producing the right mechanical resonance frequency of 6.7$~$MHz: assuming either high-stress limit (in-built stress of 60$~$MHz and neglecting the Young's modulus) or low-stress (0 in-built stress).
Besides, from Ref. \cite{cattiaux} we can produce a theoretical estimate for the Duffing parameter $\xi_m$ in Hz/m$^2$.
We obtain about $2.\times 10^{19}\,$Hz/m$^2$ for a device in the high-stress limit (a drum), and about $1.\times 10^{19}\,$Hz/m$^2$ in the low-stress case (a membrane). From the fit value quoted in Tab. \ref{tab_1} in units of Hz/phonons, we get a number in between these two numerical estimates: this validates the quantitative evaluation within $\pm 50~\%$.   

The linear coupling strength $g_0$ is defined as:
\begin{equation}
g_0  =  -G \, x_{zpf} , \label{C1}
\end{equation}
\begin{equation}
G = \frac{d \omega_{c}}{d x} = \frac{d \omega_{c}}{d C}\frac{d C}{d x} , \label{C2}
\end{equation}
with $x_{zpf}=\sqrt{\hbar/(2 m_{ef\!f}\, \omega_m)}$ the zero-point-fluctuation, defined from the mode effective mass $m_{ef\!f}$. 
For simplicity we will neglect the mode-shape here and consider two planar electrodes; as such, we will take as reference for the mode mass and spring constant calculation the center of the drum (i.e. maximum of mode shape equal to 1). Numbers are given in the caption of Tab. \ref{tab_2}.

In Eq. (\ref{C2}), we have $d \omega_{c} /d C = - \omega_{c}/(2 C_0)$ with $C_0$ the mode effective capacitance. From standard electromagnetism we write $d C/d x = + \epsilon_0 \pi R^2/d^2$, neglecting fringing effects which are small in the limit $d/R \ll 1$ ($\epsilon_0$ being the vacuum permittivity)  \cite{feynmanbook}. By definition, we take the direction of the X-axis pointing towards the fixed electrode.
Reversing the direction of the X-axis changes the sign of $g_0$ but also of $g_2$, producing an overall $(-1)^n$ in Eq. (\ref{Jn}).
This has no impact on physical quantities (such as $\gamma_{BA}$, $\delta \omega$ and $\vert \alpha_n \vert^2$): the problem at stake is invariant under a mirror symmetry.
We then obtain from Eq. (\ref{C1}) a value of about 20$~$Hz for $g_0$ (choosing $g_0>0$) taking for the cavity mode $C_0 \approx 100~$fF, which is consistent with the microwave design. 
This over-estimates $g_0$ (by about a factor of two) since in reality not all the drum electrode moves, the borders being clamped.

Expanding the plate capacitor expression in a Taylor series of $x/d$, we obtain for the cavity resonance frequency:
\begin{eqnarray}
\omega_c (x) & = & \omega_c (0) - \left[ g_0 \frac{x}{x_{zpf}}  \right. \\ \nonumber
&+& \left. \frac{g_1}{2}\left(\frac{x}{x_{zpf}}\right)^{\!2}+  \frac{g_2}{2} \left(\frac{x}{x_{zpf}}\right)^{\!3} + \cdots \right],  \label{C3}
\end{eqnarray}
at third order, where we identify:
\begin{eqnarray}
g_1 & = & g_0 \left[+2 \frac{x_{zpf}}{d} - 3 \frac{g_0}{\omega_c(0)} \right] , \\ \nonumber
g_2 & = & g_0 \left[ +2 \left( \frac{x_{zpf}}{d} \right)^{\!2} -6\frac{x_{zpf}}{d}\frac{g_0}{\omega_c(0)}  +5 \left(\frac{g_0}{\omega_c(0)} \right)^{\!2} \right] .
\end{eqnarray}
In our case, only the first terms in the above are relevant: the magnitude with respect to $g_0$ of these $g_n$ coefficients is thus fixed by $(x_{zpf}/d)^n$.
Computing numerical estimates,
we see that with the chosen value of $d$ we under-estimate $g_1$ by only about 20$~\%$, but under-estimate $\left|g_2\right|$ by a factor of 7 approximately. The sign of $g_2$ is also not captured, which shows that this crude modeling fails for high-order derivatives. 

\section{Heating and material-dependent effects}
\label{heatTLS}

\begin{figure}[t!]
		\centering\offinterlineskip
	\includegraphics[width=11.5cm]{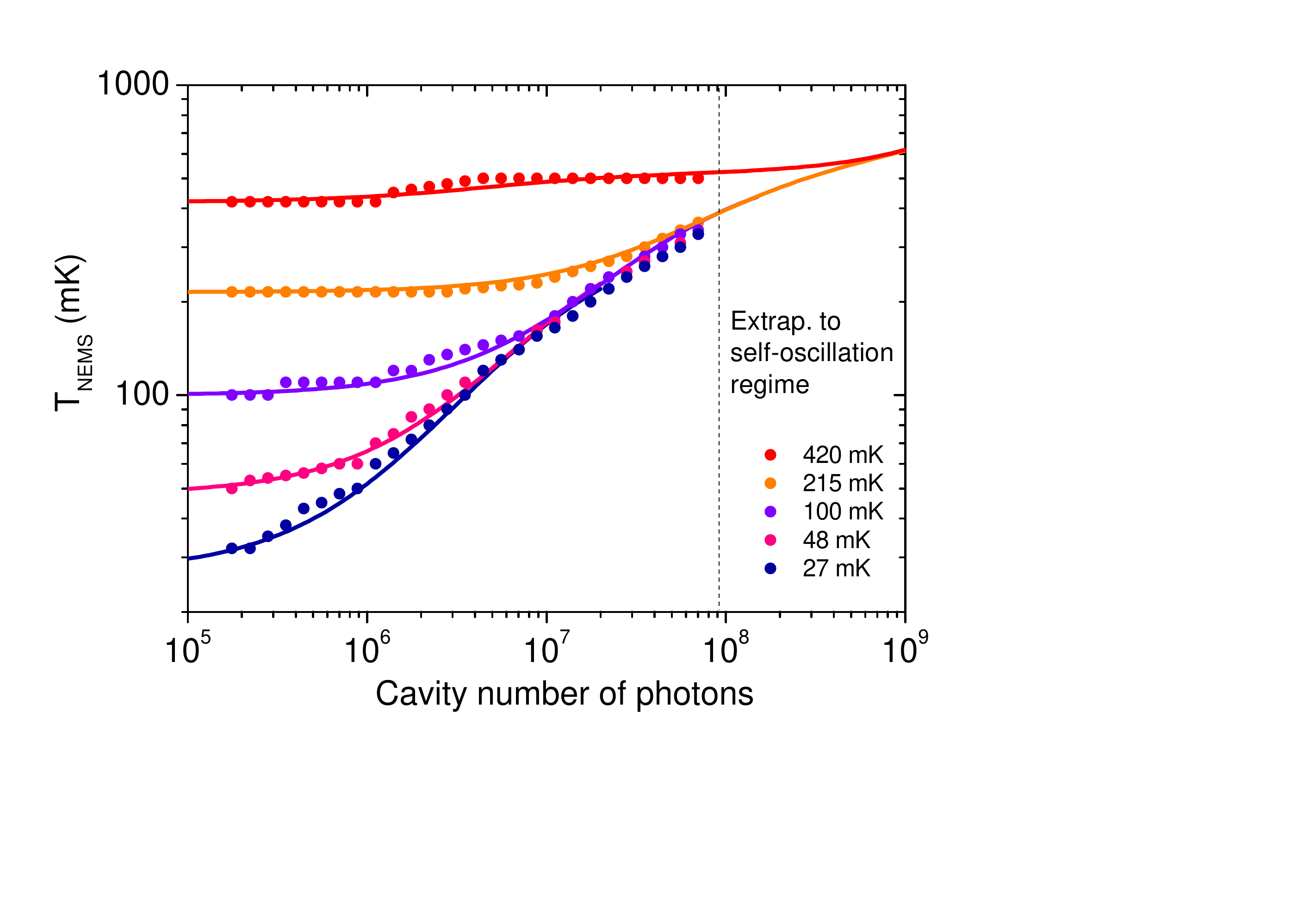}
	\vspace*{-1.5cm}
			\caption{
			(Color online) Mechanical device temperature versus applied microwave power (expressed in terms of intra-cavity photons $n_{cav}$).
			Dots are experimental data measured by red-detuned pumping, integrating the anti-Stokes power spectrum peak (see text). The curves are empirical expressions used for the extrapolation in the self-oscillating range (above $10^8$ photons). At high enough powers, all the curves collapse (the starting temperature is irrelevant compared to the added energy). The discrepancies in the numerics in the extrapolated range are smaller than $\pm 20~\%$. }
			\label{fig_6}
\end{figure}

Since we work at very large microwave powers, some material-dependent effects have to be taken into account in order to be quantitative in the fitting.
The first of these is microwave heating of dielectrics due to absorption of the radiation.
Note however that technical heating is not a fundamental effect; it can be minimized by means of phase-noise filtering and remains small for most devices with reasonable $g_0$ coupling (see e.g. Ref. \cite{sillanpaaintrique}).
In the present experiment, the setup has been kept basic and no filtering has been used. 
In order to characterize heating independently of the self-oscillating regime, we use red-detuned sideband pumping. 
As we increase the injected power $P_{in}$, we measure the area of the anti-Stokes peak. Knowing the theoretical dependence of this parameter on both $P_{in}$ and NEMS temperature $T_{NEMS}$ \cite{AKMreview}, we can recalculate $T_{NEMS}$ for each setting, see Fig. \ref{fig_6}.
This effect being local, the absorbed power has to be proportional to the intracavity field, i.e. the photon population $n_{cav}$. We can therefore extrapolate what should be the heating effects in the self-oscillating regime using the actual intracavity photon  number $\Sigma \vert \alpha_n \vert^2$. Empirical fits are shown in Fig. \ref{fig_6} (see lines). The curves merge when the heating effect dominates over the starting temperature; we therefore estimate that our extrapolation in the region of interest is accurate within $\pm 20~\%$, see Fig. \ref{fig_6}. 
Of the parameters appearing in the theory of Section \ref{theory}, the only temperature dependent ones are $\omega_c, \omega_m$ and $\gamma_m$. The mechanical damping (in Hz) is fit to measurements performed in the Brownian regime by the expression  $\gamma_m/(2 \pi) = 70.5 + 1300\,T_{NEMS}$, while the mechanical resonance frequency (in Hz) is fit by $\omega_m/(2 \pi) =6.747 \times 10^6 + 430\, \ln(T_{NEMS})$, with in both expressions $T_{NEMS}$ in K.
 
While for this sample, the mechanical element is very sensitive to heating, the microwave cavity seems to be rather insensitive. We attribute this to the fact that the cavity is much larger than the drum, and directly coupled to the substrate instead of being suspended.
However, we do measure a power dependence of the microwave resonance frequency which shifts upwards logarithmically with increasing powers. At the same time, we do not measure any change in the cavity $Q$ factor within our resolution. 
These power-dependencies of superconducting microwave resonators are commonly attributed to microscopic Two Level Systems present in the devices \cite{TLS}. Pragmatically, we take into account this effect by adding this logarithmic frequency shift to the calculation of $\omega_c$ when fitting the 3D maps: $\delta\omega_c/(2\pi)=1.8 \times 10^5 \ln(P_{in})$ in Hz.

Similarly to the Duffing effect of the mechanics, there is an equivalent nonlinearity in the microwave resonance called {\it Kerr nonlinearity}. This leads to an additional frequency shift $\propto n_{cav}$. 
This effect comes from the nonlinear behavior of the mode effective inductance $L_0$ when the current density $J$ flowing in the superconductors becomes too large \cite{paperAlessandro,Maleeva}:
\begin{equation}
L_0 (J) = L_0(0) \left[ 1 + \alpha_l \frac{J^2}{J_*^2}\right] ,
\end{equation}
with $J_*=(2/3)^{3/2} J_C$ and $J_C$ the critical current density, and $\alpha_l = L_{kin}/L_{0}$ the fraction of the total inductance of kinetic origin. For our Al film of about 100$~$nm, $\alpha_l$ should be smaller than 0.1 typically. 
 The cavity resonance frequency thus shifts as:
 \begin{equation}
 \omega_c(n_{cav}) = \omega_c(0) + \xi_c \, n_{cav} ,
 \end{equation}
 with:
 \begin{equation}
 \xi_c =  - \frac{\alpha_l \hbar \omega_c}{L_0 A^2 (2/3)^3 J_c^2},
 \end{equation}
and $A$ the cross-section of the microwave cavity strip. A crude estimate taking the bulk value for the critical current density leads to $\xi_c \approx -10^{-4}~$Rad/s, which is completely negligible.

Finally, {\it nonlinear friction} has been reported in bottom-up electro-mechanical structures made of carbon (nanotubes, graphene) \cite{bachtoldnonlindamp}.
It is taken into account by modifying the mechanical equation of motion such that $\gamma_m d x/dt \rightarrow (\gamma_m+ \gamma_2 \, x^2) d x/dt$ \cite{cross}.
In contrast for bulk top-down objects, this mechanism seems to be very small, even in cantilever devices sustaining large motion amplitudes \cite{RSIcollin}.
Experimentally, it is then rather difficult to distinguish such {\it anelastic effects} from basic Joule heating; assessing a reasonable number for the nonlinear friction coefficient $\gamma_2$ is essentially out of reach here.
On the other hand, we do have reasonable estimates for the order of magnitude of the $g_1$, $g_2$ coefficients. The quantitative fits do match these values.
Applying ``Ockham's razor'', we therefore keep the minimal set of variables necessary for the quantitative description, which is also the reason why the expansion was cut at order 3 in nonlinear coupling.

\begin{figure}[t!]
		\centering\offinterlineskip
	\hspace*{-0.5cm}
	\includegraphics[trim=3cm 11cm 0cm 0cm,clip=True,width=26.5cm]{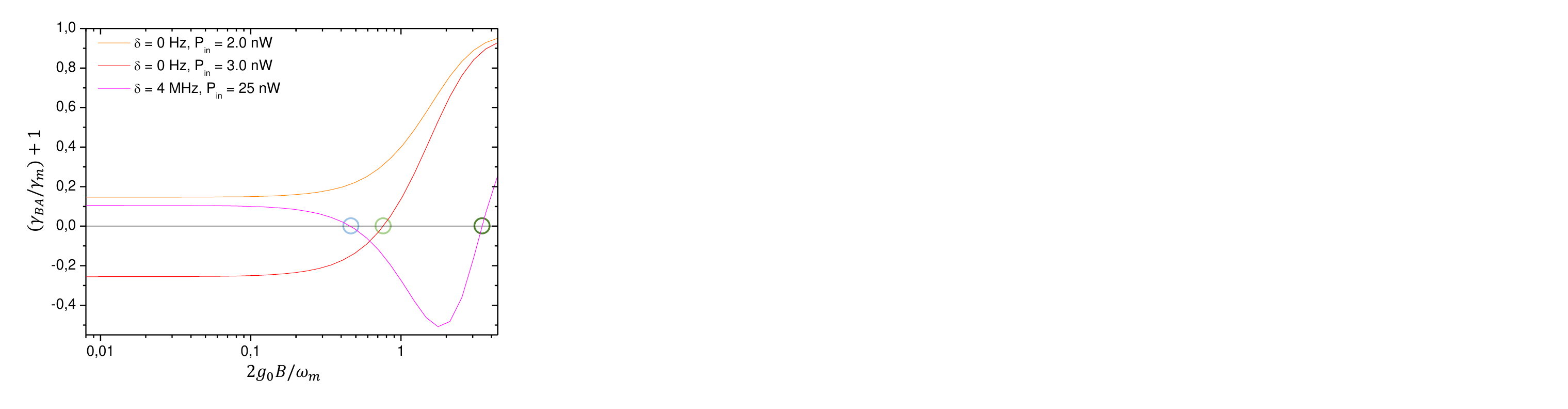}
			\caption{
			Theoretical stability curves giving $\gamma_{BA}/\gamma_m+1$ as a function of $(2g_0/\omega_m)\times B$, calculated at 3 different positions $(\delta,P_{in})$ and demonstrating the typical observed behaviors: unstable (orange line), one stable state (red line), and unstable (blue circle) plus stable states leading to hysteresis (magenta line, see text and Figs. \ref{fig_2} and \ref{fig_3}). The self-consistent value of $B$ corresponds to the (light and dark) green circles.
			}
			\label{fig_4}
\end{figure}

\section{Impact of static deflection, Duffing and Kerr nonlinearities}
\label{duffing}

\begin{figure*}[t!]		 \vspace*{-0.5cm}
\center
			 \includegraphics[trim=4.5cm 0cm 4cm 3cm,clip=false,width=15cm]{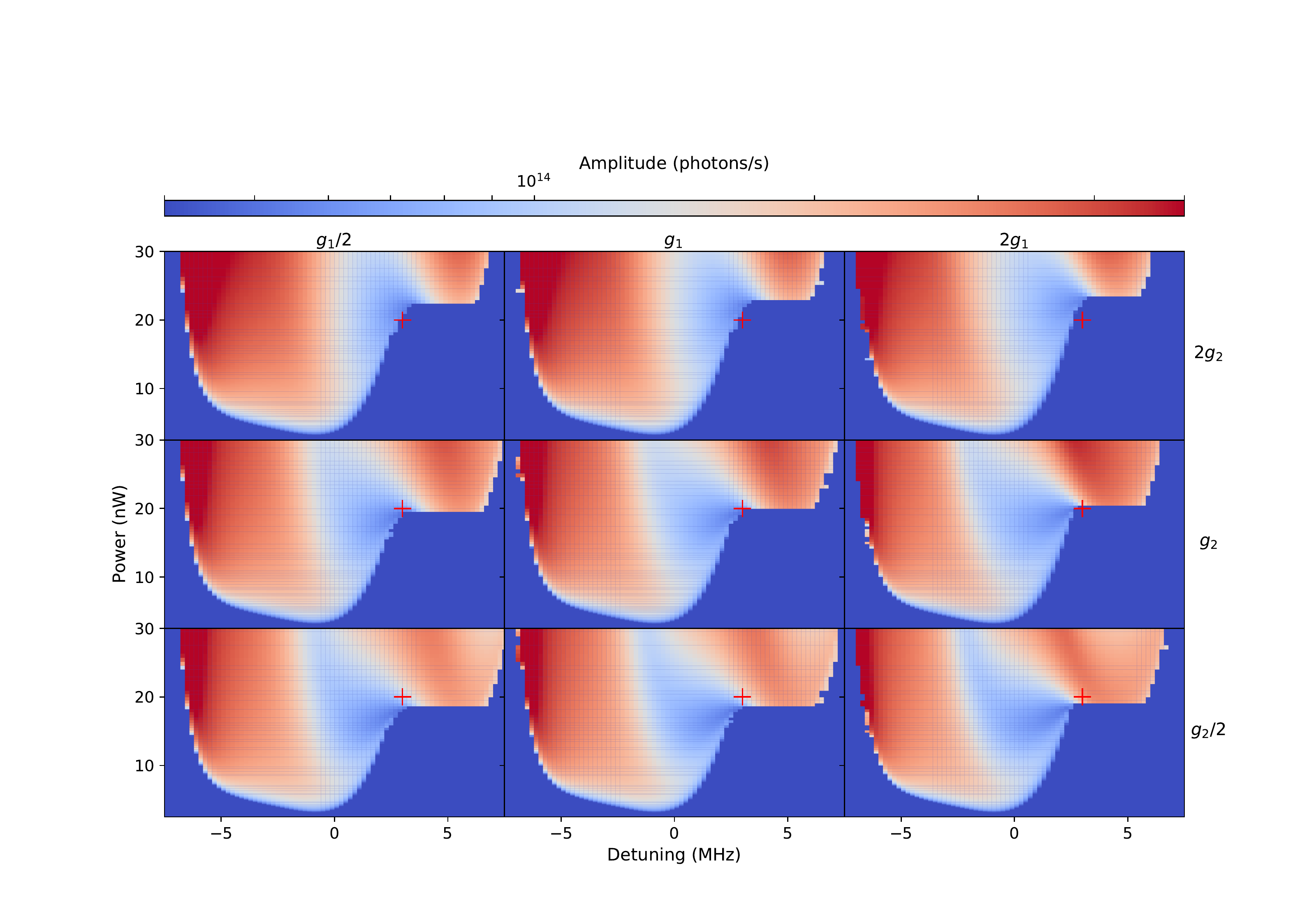}
			\vspace*{-1cm}
			\caption{
			Impact of the variation of $g_1$ and $g_2$ on the theoretical map giving the output photon flux as a function of both the detuning $\delta$ and the input pump power $P_{in}$. The colormaps are calculated taking into account all mechanical and optical shifts, with $g_0>0$. The central one is the same as in the 3D plot of Fig. \ref{fig_2}. From these graphs, one can extract the $\Delta P$, $\Delta \delta$ parameters shown in Fig. \ref{fig_g1g2}.
			The red cross marks the position of the beginning of the hysteresis for the central graph (optimal $g_1,g_2$ fit parameters). }
			\label{fig_5}
		\end{figure*}

The ansatz Eq. (\ref{betac}) introduces a static term $\beta_c$ that corresponds to a static deflection of the drum $x_c= x_{zpf} 2 \Re[\beta_c]$.
It can be deduced by solving Eq. (\ref{beta}) keeping only time-independent terms. The Duffing contribution can easily be incorporated in it.
This term remains always extremely small, and contributes only for a (tiny) cavity frequency shift $\omega_c' = \omega_c - g_1 B^2- \delta \omega_c'$:
\begin{eqnarray}
\delta \omega_c' & =& 2 g_0 \Re[\beta_c]  +2 g_1  \Re[\beta_c]^2+ 4 g_2  \Re[\beta_c]^3 \\ \nonumber
&+& 6 g_2  \Re[\beta_c] B^2  . 
\end{eqnarray} 

Since the mechanical motion can be very large (up to about 15 nm), the nonlinear stretching effect of the membrane has to be considered; this is the so-called Duffing nonlinearity, which shifts the mechanical resonance by $\xi_m B^2$.
This term can be taken into account recursively in the calculation of the stable states, Eq. (\ref{equatosolve}) see Appendix \ref{stablestates}. 
The result is that this term has only a marginal impact on the self-oscillating states definition. However, it dominates the mechanical frequency shift over the optical spring terms.
We can therefore fit $\xi_m$ on the measurement of the drum frequency, Fig. \ref{fig_3}. The obtained value is essentially temperature-independent, and given in Tab. \ref{tab_1}.

\section{Stable states computation}
\label{stablestates}

The problem is solved numerically by finding self-consistently a stable solution $B$ to Eq. (\ref{equatosolve}) for any couple ($\delta$,$P_{in}$).
These stability points correspond graphically to the intersection between the function $\gamma_{BA}(B)/\gamma_m+1$ and the X-axis (see Fig. \ref{fig_4}).
The output photon flux is thus calculated by means of  Eq. (\ref{Ndot}) injecting the found value of $B$ in Eq. (\ref{Jn}). For simplicity, one can neglect mechanical shifts which have only a marginal effect on stable states amplitudes. The procedure is then repeated over the full range of detunings $\delta$ and input pump powers $P_{in}$ in order to draw the theoretical mapping of the self-oscillating state (see green points in Fig. \ref{fig_2}). 

For small detunings, the curves are always monotonous. At low powers, there is no solution since $\gamma_{BA}(B)/\gamma_m+1>0$ (orange line in Fig. \ref{fig_4}). Increasing the power brings eventually the curve below the X-axis, creating a single intersection  $\gamma_{BA}(B)/\gamma_m+1=0$ (green circle on the red curve). Fluctuations at small $B$ can thus trigger the self-oscillating state as the curve smoothly goes below $Y=0$.

For large positive detunings, there is a range at (large) powers where the curve displays {\it two intersections} (see magenta line in Fig. \ref{fig_4}). For the low-$B$ valued one (blue circle), the slope is negative which means that the state displays anti-damping: it is unstable.
On the other hand, for the high-$B$ solution the derivative is positive, which means that the state is stable (dark green circle).

However, this state is at very large amplitudes $B$, and {\it has not been} created by a smooth crossing of the X-axis from the whole curve, starting at the lowest $B \approx 0$: this means that it can be triggered only if one comes already from high amplitude states, and not from thermal motion. This is exactly the hysteretic behavior that is seen in Fig. \ref{fig_2}, sweeping the detuning $\delta$ upwards at constant power, and increasing the power from typically 2 nW to 30 nW. The same is true, sweeping the power downwards from the high-$B$ state at fixed detuning.

The graphs in Fig. \ref{fig_4} are obtained with $g_1=+10^{-7}\, g_0$ and $g_2=- 10^{-13}\, g_0$ (with $g_0 >0$).
The numerical calculation can be performed with the static deflection and the Duffing term taken into account, see Appendix \ref{duffing}. The results are essentially identical.
The fitting routine is explained in the next Appendix.

A similar nonlinear effect exists for the cavity: this is the Kerr effect already discussed in Appendix \ref{heatTLS}.
At first order, this term shifts the position of the resonance by a quantity $\xi_c \vert \alpha \vert^2$. The expected value for $\xi_c$ being very small, we can simply completely neglect any nonlinear effect of that sort.

\section{3D Fitting procedure}
\label{fitting}
		
Measurements of the photon flux are compared to the theoretical computation in Fig. \ref{fig_2}.
The two parameters $g_1$ and $g_2$ affect the shape of the numerical $(\delta,P_{in})$ colormap in different ways: 
we demonstrate this in Fig. \ref{fig_5} varying them in a dichotomic process (multiplying or dividing the optimal values by 2).
By increasing  $ g_1 $ the self-oscillating region is getting more narrow in the $\delta$ direction, while  increasing $\vert g_2 \vert$ up-shifts in power the starting line of the bistable region. This is discussed in the core of the paper with the parameters $\Delta P$, $\Delta \delta$, see Fig. \ref{fig_g1g2}. The optimal values match the experimental findings: $\Delta P \approx 16~$nW ($\pm 10~\%$), $\Delta \delta \approx 3.5~$MHz ($\pm 200~$kHz).

We can therefore choose the red cross position in Fig. \ref{fig_5} (central graph, optimal $g_1$ and $g_2$) as a good marker for fitting these $g_1$ and $g_2$ parameters (equivalent of Fig. \ref{fig_nononlin}, but $ g_1 = g_2=0 $ value). 
Since the NEMS heats with applied power, this also defines the actual temperature at which the fit is essentially performed. This is summarized in Tab. \ref{tab_1}, with error bars estimated for the coupling nonlinear parameters to be about a factor of 2. Fits of the mechanical frequency shifts are discussed in Appendix \ref{duffing}; Fig. \ref{fig_3} is essentially an image of the amplitude of motion squared $x^2$ (or equivalently $B^2$). Note that the quality of the agreement between experiment and theory in this graph also imposes strong constraints on the $(g_1,g_2)$ couple. This is also the case of the overall shape of the photon flux maps, Fig. \ref{fig_2}.

%
%


\end{document}